\documentclass[british,english,journal]{IEEEtran}
\usepackage[latin9]{inputenc}
\usepackage{amsmath}
\usepackage{amsthm}
\usepackage{amssymb}
\usepackage{graphicx}
\usepackage{esint}
\usepackage{cases}
\makeatletter
\theoremstyle{plain}

\theoremstyle{definition}

\usepackage{cite}

\makeatletter
\newcommand{\unfootnote}[1]{
  \renewcommand{\@makefnmark}{}
  \footnotetext{#1}
  \renewcommand{\@makefnmark}{\mbox{$^{\@thefnmark}$}}
}
\makeatother

\usepackage{cite}

\usepackage{subfigure}

\usepackage{cite}

\AtBeginDocument{
  
}

\makeatother

\usepackage{babel}
\addto\captionsbritish{\renewcommand{\examplename}{Example}}
\addto\captionsbritish{\renewcommand{\theoremname}{Theorem}}
\addto\captionsenglish{\renewcommand{\examplename}{Example}}
\addto\captionsenglish{\renewcommand{\theoremname}{Theorem}}
\providecommand{\examplename}{Example}
\providecommand{\theoremname}{Theorem}

\begin{document}

\title{Performance Analysis of the Raft Consensus Algorithm for Private Blockchains}

\author{Dongyan Huang, Xiaoli Ma, \IEEEmembership{Fellow, IEEE,} and Shengli Zhang, \IEEEmembership{Senior Member, IEEE}
\thanks{Manuscript received June 27, 2018.}\thanks{D. Huang is with the College of Information Engineering at Shenzhen University, \IEEEauthorblockA{Email: huangdongyan-gua@163.com.} X. Ma is with the School of Electrical and Computer and Engr. at Georgia Institute of Technology, \IEEEauthorblockA{Email: xiaoli@gatech.edu.} Part of the work was supported by QuarkChain Foundation LTD where Dr. Ma was working as a consultant. S. Zhang is with the College of Information Engineering at Shenzhen University, \IEEEauthorblockA{Email: zsl@szu.edu.cn.}}}

\maketitle
\begin{abstract}
Consensus is one of the key problems in blockchains. There are many articles analyzing the performance of threat models for blockchains. But the network stability seems lack of attention, which in fact affects the blockchain performance. This paper studies the performance of a well adopted consensus algorithm, Raft, in networks with non-negligible packet loss rate. In particular, we propose a simple but accurate analytical model to analyze the distributed network split probability. At a given time, we explicitly present the network split probability as a function of the network size, the packet loss rate, and the election timeout period.
To validate our analysis, we implement a Raft simulator and the simulation results coincide with the analytical results.
With the proposed model, one can predict the network split time and probability in theory and optimize the parameters in Raft consensus algorithm.
\end{abstract}



\selectlanguage{english}%

\section{Introduction}

\IEEEPARstart{B}{lockchain} technology, which was firstly coined in Bitcoin\cite{IEEEhowto:Nakamoto} by Satoshi Nakamoto in 2008, has received extensive attentions recently. A blockchain is an encrypted, distributed database/transaction system where all the peers share information in a decentralized and secure manner. Due to its key characteristics as decentralization, immutability, anonymity and auditability, blockchain becomes a promising technology for many kinds of assets transfer and point-to-point (P2P) transaction. Recently, blockchain-based applications are springing up, covering numerous fields including financial services\cite{IEEEhowto:Peters, IEEEhowto:Kosba, IEEEhowto:Akins}, Internet of Things (IoT)\cite{IEEEhowto:Zhang}, reputation systems\cite{IEEEhowto:Sharples}, and so on.

Since the essence of blockchain is a distributed system, consensus algorithms play a crucial role in maintaining the safety and efficiency of blockchains. A consensus algorithm for distributed systems is to figure out the coordination among multiple nodes, i.e., how to come to agreement if there are multiple nodes. Several consensus algorithms have been proposed, e.g., Proof-of-Work (PoW)\cite{IEEEhowto:Nakamoto}, Proof-of-Stake (PoS)\cite{IEEEhowto:King}\cite{IEEEhowto:Nxtwiki}, Delegate Proof-of-Stake (DPoS)\cite{IEEEhowto:https}, Practical Byzantine Fault-Tolerant (PBFT)\cite{IEEEhowto:Castro}, Paxos\cite{IEEEhowto:Lamport} and Raft\cite{IEEEhowto:Ongaro}. Among them, PoW and PoS algorithms have a good support for safety, fault tolerance and scalability of a blockchain. Thus, PoW and PoS are common choices of public blockchains, in which any one can join the network and there are no trust relationships among the nodes. However, PoW and PoS have slow speed of transaction confirmation, which limits its applications to those requiring high confirmation speed. In a consortium/private blockchain network, all participants are whitelisted and bounded by strict contractual obligations to behave ``correctly", and hence more efficient consensus algorithms such as PBFT and Raft are more appropriate choices. Consortium/private blockchains could be applied into many business applications. For example, Hyperledger\cite{IEEEhowto:Hyperledger} is developing business consortium blockchain frameworks. Ethereum has also provided tools for building consortium blockchains\cite{IEEEhowto:Consortium}. Raft algorithm is considered as a consensus algorithm for private blockchains\cite{IEEEhowto:Mingxiao}, which is applied to more ad hoc networks such as the intranet. Furthermore, several hybrid consensus protocols have been proposed to improve consensus efficiency without undermining scalability. For example, Zilliqa\cite{IEEEhowto:ZILLIQA} proposed PoW to elect directory service (DS) committee nodes, and then the DS committee has to run PBFT consensus protocol on the transaction block.

Compared with PBFT and Paxos, Raft algorithm has high efficiency and simplicity and it has been widely adopted in the distributed systems. Raft is a leader-based algorithm, which uses leader election as an essential part for the consensus protocol. Ledger entries in Raft-based system flow in only one direction from the leader to other servers. Paxos and its improved algorithms do not take leader election as an essential part of the consensus protocol, which can balance load well among nodes because any node may commit commands\cite{IEEEhowto:Lamport}\cite{IEEEhowto:MORARU}. However, Paxos' architecture requires complex changes to support practical systems. Raft achieves the same safety performance as Paxos and is more convenient in engineering implementation and understanding. Raft consensus algorithm cannot tolerate malicious nodes and can tolerate up to 50\% nodes of crash fault. For private blockchains, nodes are verified members. Hence, it is more important to solve the crash faults than Byzantine faults for private blockchains.

Network is called split when more than half of nodes are out of current leader's control. Failure of node and communication interruption caused by packet loss are the main reasons of network split. If network split occurs, the blockchain network with Raft consensus algorithm would re-start a new leader election process. Meanwhile, the blockchain network stops accepting new transaction, i.e., the blockchain network becomes unavailable. Obviously, consensus efficiency of blockchain is degraded tremendously if network split occurs frequently. The existing works on blockchain mainly focus on designing algorithms or optimizing performance or safety certification, but lack of a theoretical analysis of network split. The impact of packet loss rate on network split is rarely considered. In fact, packet loss rate plays an important role on network split.

In this paper, we concentrate on analyzing the network split probability of the Raft algorithm. We provide a simple model that accounts for protocol details, and allows to compute the performance of distributed networks.
The main contributions of this paper are given as follows:
1) a simple analytical model is developed;
2) the network performance in normal conditions is derived; and
3) we explore the parameters' (such as packet loss rate, period of election timeout and size of network) impacts on network split performance.

\section{Review of Raft algorithm}
\begin{figure}
\begin{centering}
\includegraphics[scale=0.4]{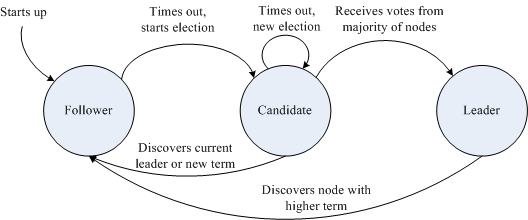}
\end{centering}
\centering{}\caption{State transition model for the Raft algorithm}
\label{fig_sim0}
\end{figure}

This section briefly summarizes the Raft algorithm. A more complete and detailed description of Raft refers to \cite{IEEEhowto:Ongaro}. Raft is a consensus algorithm for managing a replicated ledger at every node. At any given time, each node is in one of the three states:
leader, follower, or candidate. Raft algorithm divides time into terms with finite duration. Terms are numbered with consecutive integers. Each term begins with an election, in which one or more candidates
attempt to become a leader. If a candidate wins the election, then it serves as a leader for the rest of the term. The state transition is shown in Fig. \ref{fig_sim0}.  All nodes start from the follower state. If a follower does not hear from the leader for a certain period of time, then it becomes a candidate. The candidate then requests votes from other nodes to become a leader. Other nodes will reply to the vote request. If the candidate gets votes from a majority of the nodes, it will become a leader. This process is called Leader Election. Specifically, if a follower receives a heartbeat within the minimum election timeout of hearing from a current leader, it does not grant its vote to the candidate. This helps maximizing the duration of a leader to keep working and avoiding frequent disruptions from some isolated/removed nodes. \footnote{These nodes will not receive
heartbeats, so they will time out and start new elections.}

In normal operation of Raft, there is exactly one leader and all the other nodes are followers.
The leader periodically sends out heartbeats to all followers in order to maintain its authority. All transactions during this term go through the leader. Each transaction is added as an entry in the node's ledger. Specifically, the leader first replicates the received new transaction to the followers. At this time, the entry is still uncommitted and stays in a volatile state. After the leader receives feedbacks from a majority of followers that have written the entry, the leader notifies the followers that this entry is committed. This process is called Ledger Replication.

In Raft algorithm, there are several timeout settings. One of them controls the election process. The election timeout is the amount of time that a follower needs to wait to become a candidate. The election time counter is decreased as long as the follower receives no heartbeat. The follower transfers to candidate state when the election time reaches zero. The election time counter resets to a random value when the follower receives a heartbeat from a leader. The random election timers in Raft help to reduce the probability that several followers transfer to candidates simultaneously.



\section{system model}
In this section, we focus on the analysis for network split probability. Consider a distributed network with $N$ nodes and $N$ is odd. \footnote{When $N$ is even, the algorithm still works and our claims still hold. Here for the convenience of analysis, we consider odd $N$. So that $N-1$ is even.}The way of message exchanging among nodes is according to the Raft algorithm.
To ensure the system efficiency, the interval between heartbeats is much less than election timeout. The average time between failures of a single
node is much larger than election timeout.

A network split happens if more than half of nodes are out of the current leader's control.
We try to inverstigate the relationship between the parameters (such as timeout counter, packet loss rate and network size) and network split probability. First, the process of one follower node transfers to a candidate is modeled as an absorbing Markov chain. Then the network spilt probability is derived based on the model. Finally, the expected time for a node transferring from the follower to the candidate and the expected number of received heartbeats for one node are derived.

In the following analysis, within one term the initial state is defined as after a successful received heartbeat, the follower's election counter is reset. Suppose that there is a leader and $N-1$ followers. We assume that the communication delay is much less than the heartbeat interval. One heartbeat means one step in the Markov chain.

\subsection{Network model}
\

Define the packet loss probability as $p$, and suppose that $p$ is a constant value for a given network. Denote the timeout value for each round of election as $E_{t}$, which is initially uniformly chosen from the range $[a,b]$. The interval between two heartbeats is $h$. A discrete and integer time scale is adopted. Thus, if a follower fails to receive $K=\lfloor E_{t}/h\rfloor$ heartbeats consecutively, then it assumes there is no viable leader and transitions to candidate state to start an election. Noted that $K\in\{K_{1},K_{2},\cdots,K_{r}\}$, and $K$ is uniformly chosen from the set $\{K_{1},K_{2},\cdots,K_{r}\}$, where $K_{1}=\lfloor a/h\rfloor$ and $K_{r}=\lfloor b/h\rfloor$. In the following analysis, $K$ denotes as the maximum number of heartbeats for an election counter to timeout.


Let $g(n)$ be the stochastic process representing the stage status $\{1,2,\cdots,r\}$ of a given node at time $n$. Let $b(n)$ be the stochastic process representing the left steps of election time counter for the node at time $n$. Once the independence between $g(n)$ and $b(n)$ is assumed, we can model it as a two-dimensional process $\{g(n),b(n)\}$.

We adopt the short notation: $P\{i,k_{i}-1\mid i,k_{i}\}=P\{g(n+1)=i,b(n+1)=k_{i}-1\mid g(n)=i,b(n)=k_{i}\}$.
In this Markov chain, the only non-null one-step transition probabilities are
\begin{numcases}{}
P\{i,k_{i}-1\mid i,k_{i}\} =p \label{equ_2-1}\\
P\{i,K_{i}\mid j,k_{j}\}=(1-p)/r \label{equ_2-2}\\
P\{i,0\mid i,0\}=1, \label{equ_2-3}
\end{numcases}
where $i,j={1,2,\cdots,r}$ and $k_{i}\in\{1,\cdots,K_{i}\}$. Eq. (\ref{equ_2-1}) relies on the fact that the follower fails to receive a heartbeat from the current leader and its election time counter is decreased by 1. Eq. (\ref{equ_2-2}) shows the fact that the follower receives a heartbeat and reset the election time counter. Eq. (\ref{equ_2-3}) shows that once the election time counter reaches zero, the follower transitions to the candidate state.

Denote $\{i,0\}$ as the absorbing state. Since $i={1,2,\cdots,r}$, there are $r$ absorbing states. Denote the other states except state $\{i,0\}$ in state space of $\{g(n),b(n)\}$ as transient states. There are $t$ transient states, where $t=\sum \limits_{i=1}^{r}K_{i}$. Let us order the states such that the first $t$ states are transient and the
last $r$ states are absorbing. The transition matrix has the following canonical form
\begin{equation}
 \textbf{P}=\left(
  \begin{array}{cc}
    \textbf{Q} & \textbf{R} \\
    \textbf{0} & \textbf{I} \\
  \end{array}
\right), \label{equ_2-300}
\end{equation}
where $\textbf{Q}$ is a $t\times t$ matrix, $\textbf{R}$ is a nonzero $t\times r$ matrix, the entries of $\textbf{R}$ are the probabilities the transient states transfer to absorbing states, $\textbf{0}$ is an $r\times t$ zero matrix, and $\textbf{I}$ is an $r\times r$ identity matrix. Specifically, the entry $q_{ij}$ of $\textbf{Q}$ is defined as the transition probability from transient state $s_i$ to transient state $s_j$, the entry $r_{mn}$ of $\textbf{R}$ is defined as the transition probability from transient state $s_m$ to absorbing state $s_n$.

When $K_{r}-K_{1}\ll K_{1}$ or $b-a<h$, the election timeout only has value one. Thus, $r=1$, $t=K$ and then the only non-null one-step transition probabilities in Eqs. (\ref{equ_2-1})-(\ref{equ_2-3}) can be simplified as follows:
\begin{numcases}{}
P\{k-1\mid k\} =p, \label{equ_2-4}\\
P\{K\mid k\}=1-p,  \label{equ_2-5}\\
P\{0\mid 0\}=1, \label{equ_2-6}
\end{numcases}
where $k\in\{1,\cdots,K\}$. Thus, the transition matrix $\textbf{P}$ becomes a $(K+1)\times(K+1)$ matrix as

\begin{equation}
 \textbf{P}=\left(
  \begin{array}{cccc}
    1-p    & p &  & \mathbf{0}\\
    \vdots &   & \ddots & \\
    1-p    &  \mathbf{0} &  & p \\
    \mathbf{0} &  &  &1\\
  \end{array}
\right). \label{equ_2-100}
\end{equation}

At the $n$th step, the transition matrix is $\textbf{P}^{n}$ and
the entry $p^{(n)}_{ij}$ of the matrix $\textbf{P}^{n}$ is the probability of
being in the state $s_j$ from the state $s_i$.

For simplicity, suppose that election timeout counter has a fixed value $K$ in the following analysis. It is straightforward to extend the analytical results to the case in which election timeout is a random value, i.e., $r>1$.

\subsubsection{Network split probability}
\

Consider a network with one leader and $N-1$ followers. Note that when more than half of $N$ nodes become candidates, the leader will not be qualified and thus the network will split. For network split probability, we have the following proposition.

\textbf{Proposition 1:} For a network with $N$ nodes, the transition matrix is given in Eq. (\ref{equ_2-100}). Then, the probability of network split before the $n$th step is given by
\begin{equation}
p_{n} = 1- \sum_{m=0}^{\lfloor\frac{N}{2}\rfloor } \binom{N-1}{m}\left (p_{1(K+1)}^{(n)}\right )^{m}\left (1-p_{1(K+1)}^{(n)}\right )^{N-1-m} \label{equ_2-19}
\end{equation}
where $p_{1(K+1)}^{(n)}$ is the $(1,K+1)$th entry of the matrix $\textbf{P}^{n}$.

\textbf{Proof:}
The entry $p_{i(K+1)}^{(n)}$ of the matrix $\textbf{P}^{n}$ is the probability of being absorbed before the $n$th step, when the chain is started from state $s_i$. Thus,
$p_{1(K+1)}^{(n)}$ is the probability of the follower starts from the initial state and transits to candidate state before the $n$th step.

Denote $Y_{n}$ being the number of nodes which transit to candidate state before the $n$th step given that all nodes start from the initial state. Thus, $P\{Y_n=m\}$ is the probability that $m$ nodes become candidates before the $n$th step. Suppose that all nodes are independent. Therefore, $Y_{n}$ is a binomial distribution random variable with the form:

\begin{equation}
P\{Y_n = m\} = \binom{N-1}{m}\left (p_{1(K+1)}^{(n)}\right )^{m}\left (1-p_{1(K+1)}^{(n)}\right )^{N-1-m}. \label{equ_2-17}
\end{equation}

Therefore, $P\{Y_n \geq \left \lfloor \frac{N}{2} \right \rfloor +1 \}$ is the probability that more than half of followers become candidates before the $n$th step. We have
\begin{equation}
p_{n}=P \left \{Y_n \geq  \left \lfloor \frac{N}{2} \right \rfloor +1 \right \} = 1- \sum_{m=0}^{\left \lfloor\frac{N}{2}\right \rfloor }P\{Y_n=m\}. \label{equ_2-190}
\end{equation}

Thus, Proposition 1 is proved.\ \ \ \ \ \ \ \ \ \ \ \ \ \ \ \ \ \ \ \ \ \ \ \ \ \ \ \ \ \ \ \ \ \ \ \ \ \ \ \ \ \ \ \ \ \ \ \ \ \ \ \ \ \ \ \ \ \ \ \ \ \ \ \ \ \ \ \ \ \ \ \ \ \ \ \ \ \ \ \ \ \ \ \ \ \ \ \ \ \ \ \ \ \ \ \ \ \ \ \ \ \ \ \ \ \ \ \ \ \ \ \ \ \ $\blacksquare$

Based on Proposition 1, we derive the following properties of network split probability.

\textbf{Property 1}: The transition probability at the $n$-th step $p_{1(K+1)}^{(n)}$ has the following form:
\begin{equation}
p_{1(K+1)}^{(n)}=\left\{ \begin{array}{l l l l}
 0,  &  \mbox{if $n<K$};\\
 p^{K},  &  \mbox{if $n=K$};\\
 p_{1(K+1)}^{(n-1)}+\left (1-p_{1(K+1)}^{(n-K-1)}\right )(1-p)p^{K},  &  \mbox{if $n>K$};\\
\end{array} \right.
\label{equ_2-300}
\end{equation}

\textbf{Proof:} According to the transition matrix $\mathbf{P}$ shown in Eq. (\ref{equ_2-100}), $p_{1(K+1)}^{(n)}$ has the following forms when $n<K$ and $n=K$.

If $n<K$, then a follower cannot transit to the candidate state. Therefore,
\begin{equation}
p_{1(K+1)}^{(n)}=0.
\label{equ_2-301}
\end{equation}

If $n=K$, by calculating $\textbf{P}^{n}$, we obtain
\begin{equation}
p_{1(K+1)}^{(n)}=p^{K}.
\label{equ_2-302}
\end{equation}

In the following analysis, we derive the form of $p_{1(K+1)}^{(n)}$ when $n>K$.
According to $\textbf{P}$ shown in Eq. (\ref{equ_2-100}), we have
\begin{equation}
p_{1(K+1)}^{(n)} = p_{1(K+1)}^{(n-1)}+p\cdot p_{1K}^{(n-1)},
\label{equ_2-303}
\end{equation}

\begin{equation}
p_{1i}^{(n)} = p\cdot p_{1(i-1)}^{(n-1)}, \ \ \ \  \ \forall i\neq 1\ \mathrm{ and} \  i\neq K+1,
\label{equ_2-304}
\end{equation}
and
\begin{equation}
p_{11}^{(n)} = (1-p)\cdot \sum_{i=1}^{K}p_{1i}^{(n-1)}.
\label{equ_2-305}
\end{equation}

From Eq. (\ref{equ_2-304}), we obtain
\begin{equation}
p_{1K}^{(n)} = p^{K-1}\cdot p_{11}^{(n-K+1)}.
\label{equ_2-307}
\end{equation}

Since the row sums of transition matrix $\mathbf{P}$ are equal to one, we obtain
\begin{equation}
\sum_{i=1}^{K}p_{1i}^{(n-1)}=1-p_{1(K+1)}^{(n-1)}.
\label{equ_2-306}
\end{equation}

By combining Eq. (\ref{equ_2-306}) with Eq. (\ref{equ_2-305}), we have
\begin{equation}
p_{11}^{(n)} =(1-p)\left(1-p_{1(K+1)}^{(n-1)} \right).
\label{equ_2-308}
\end{equation}

By combining Eq. (\ref{equ_2-307}) and Eq. (\ref{equ_2-308}) to Eq. (\ref{equ_2-303}), we obtain
\begin{equation}
p_{1(K+1)}^{(n)}=p_{1(K+1)}^{(n-1)}+\left (1-p_{1(K+1)}^{(n-K-1)}\right )(1-p)p^{K}.
\label{equ_2-21}
\end{equation}

Thus, Property 1 is proved.\ \ \ \ \ \ \ \ \ \ \ \ \ \ \ \ \ \ \ \ \ \ \ \ \ \ \ \ \ \ \ \ \ \ \ \ \ \ \ \ \ \ \ \ \ \ \ \ \ \ \ \ \ \ \ \ \ \ \ \ \ \ \ \ \ \ \ \ \ \ \ \ \ \ \ \ \ \ \ \ \ \ \ \ \ \ \ \ \ \ \ \ \ \ \ \ \ \ \ \ \ \ \ \ \ \ \ \ \ \ \ \ $\blacksquare$

\textbf{Property 2}: When the network size $N\rightarrow\infty$ and $p_{1(K+1)}^{(n)}\rightarrow 0$, the number of nodes transferring to candidate state before the $n$th step $Y_{n}$ is approximated by Poisson distribution random variable, and $Y_{n}\sim \mathcal{P}\left ((N-1)p_{1(K+1)}^{(n)}\right )$. The probability of network split before the $n$th step is given by

\begin{equation}
p_{n} = 1- \sum_{m=0}^{\left \lfloor\frac{N}{2} \right \rfloor }e^{(N-1)p_{1(K+1)}^{(n)}} \frac{\left ((N-1)p_{1(K+1)}^{(n)}\right )^{m}}{m!}.
\label{equ_2-21}
\end{equation}

\subsubsection{The average number of replies}
\

Since $Y_{n}$ is a binomial distribution random variable, given the network size $N$, the expected value of the number of candidates at the $n$th step is
\begin{equation}
N_{C}^{(n)}=(N-1)p_{1(K+1)}^{(n)}.\label{equ_2-18}
\end{equation}

The expected value of the number of followers at the $n$th step is

\begin{equation}
N_{f}^{(n)} = (N-1)(1-p_{1(K+1)}^{(n)}). \label{equ_2-20}
\end{equation}

Thus, the average number of replies collected by the leader in the $n$th step is

\begin{equation}
E(N_{reply}) = N_{f}^{(n)}.\label{equ_2-20}
\end{equation}

\subsubsection{The expected number of received heartbeats for a follower}
\

\textbf{Proposition 2:} Given that the follower starts from the initial state, then the expected number of received heartbeats before it transfers to candidate state is $n_{11}$, where $n_{11}$ is the first entry of matrix $\textbf{N}=(\textbf{I}-\textbf{Q})^{-1}$.

\textbf{Proof:} According to the theorem of absorbing Markov chain in \cite{IEEEhowto:Markov}, the entry of matrix $\textbf{N}=(\textbf{I}-\textbf{Q})^{-1}$ is the expected number of times the chain is in state $s_j$, given that it starts in state $s_i$. Detailed proof is given as follows.

Since
\begin{equation}
 \textbf{P}=\left(
  \begin{array}{cc}
    \textbf{Q} & \textbf{R} \\
    \textbf{0} & \textbf{I} \\
  \end{array}
\right),\nonumber
\end{equation}

we have
\begin{equation}
 \textbf{P}^{n}=\left(
  \begin{array}{cc}
    \textbf{Q}^{n} & (\textbf{Q}^{n-1}+\textbf{Q}^{n-2}+\cdots+\textbf{Q}+\textbf{I})\textbf{R} \\
    \textbf{0} & \textbf{I} \\
  \end{array}
\right).\\
\end{equation}


Note that
\begin{equation}
(\textbf{I}-\textbf{Q})(\textbf{I}+\textbf{Q}+\textbf{Q}^{2}+\cdots+\textbf{Q}^{n-1})=\textbf{I}-\textbf{Q}^{n}. \label{equ_2-7}
\end{equation}

In Appendix, we prove that the absolute values of the eigenvalues of $\textbf{Q}$ are all strictly less 1. Thus, $\textbf{I}-\textbf{Q}$ is invertible. Define $\textbf{N}=(\textbf{I}-\textbf{Q})^{-1}$. Multiplying both sides of Eq. (\ref{equ_2-7}) by $\textbf{N}$ gives
\begin{equation}
\textbf{I}+\textbf{Q}+\textbf{Q}^{2}+\cdots+\textbf{Q}^{n-1}=\textbf{N}(\textbf{I}-\textbf{Q}^{n}). \nonumber
\end{equation}

Thus, we have
\begin{equation}
\textbf{P}^{n}=\left(
  \begin{array}{cc}
    \textbf{Q}^{n} & (\textbf{I}-\textbf{Q})^{-1}(\textbf{I}-\textbf{Q}^{n})\textbf{R} \\
    \textbf{0} & \textbf{I} \\
  \end{array}
\right).
\end{equation}

In the appendix, we also prove that when $n$ goes to infinity, $\textbf{Q}^{n}$ goes to $\textbf{0}$. Therefore, when $n$ goes to infinity,
\begin{equation}
\textbf{N}=\textbf{I}+\textbf{Q}+\textbf{Q}^{2}+\cdots, \label{equ_2-9}
\end{equation}
and thus
\begin{equation}
n_{ij}=q_{ij}^{(0)}+q_{ij}^{(1)}+q_{ij}^{(2)}+\cdots, \label{equ_2-10}
\end{equation}
where $q_{ij}^{(k)}$ is defined as the $(i,j)$th entry of $\textbf{Q}^{k}$.

Let $X^{(k)}$ be a random variable, and
\begin{equation}
X^{(k)}=\left\{ \begin{array}{l l}
\mbox 1  &  \mbox{if the chain is in state $s_j$ after $k$ steps};\\
\mbox 0  &  \mbox{otherwise}.\\
\end{array} \right.
\label{equ_2-11}
\end{equation}

According to $\mathbf{P}^{k}$, we have
\begin{equation}
P(X^{(k)}=1)=p_{ij}^{(k)}=q_{ij}^{(k)},  \ \ \ \    i,j=1,\cdots,K, \label{equ_2-12}
\end{equation}

and

\begin{equation}
P(X^{(k)}=0)=1-q_{ij}^{(k)}. \label{equ_2-13}
\end{equation}
These equations hold when $k=0$ since $\textbf{Q}^{0}=\textbf{I}$. Therefore, $E(X^{(k)})=q_{ij}^{(k)}$.

The expected number of times the chain is in state $s_j$ in the first $n$ steps, given that it starts from state $s_i$,
\begin{equation}
E(X^{(0)}+X^{(1)}+\cdots+X^{(n)})=q_{ij}^{(0)}+q_{ij}^{(1)}+\cdots+q_{ij}^{(n)} \label{equ_2-14}
\end{equation}

Letting $n$ goes to infinity, we have
\begin{equation}
E(X^{(0)}+X^{(1)}+\cdots)=q_{ij}^{(0)}+q_{ij}^{(1)}+\cdots \label{equ_2-15}
\end{equation}

By comparing Eq. (\ref{equ_2-10}) with Eq. (\ref{equ_2-15}), we obtain that the entry of matrix $\textbf{N}$ is the expected number of times the chain is in state $s_j$, given that it starts from state $s_i$. Denoting $s_1$ as the initial state, $n_{11}$ is the expected number of received heartbeats before a follower transfers to candidate state when it starts from the initial state.\ \ \ \ \ \ \ \ \ \ \ \ \ \ \ \ \ \ \ \ \ \ \ \ \ \ \ \ \ \ \ \ \ \ \ \ \ \ \ \ \ \ \ \ \ \ \ \ \ \ \ \ \ \ \ \ \ \ \ \ \ \ \ \ \ \ \ \ \ \ \ \ \ \ \ \ \ \ \ \ \ \ \ \ \ \ \ \ \ \ \ \ \ \ \ \ $\blacksquare$

\subsubsection{Time to transition to candidate}
\

\textbf{Proposition 3:} Suppose that a follower starts from the initial state, the expected time for this follower to transition to candidate state is given as

\begin{equation}
t_{c}= \sum_{j=1}^{t}n_{1j}, \label{equ_2-22}
\end{equation}

where $n_{1j}$ is the $(1,j)$th entry of matrix $\textbf{N}$.

\textbf{Proof:}
According to Proposition 2, the entry $n_{ij}$ of $\textbf{N}$ gives the expected number
of times that the follower is in the transient state $s_j$ if it is started from the transient
state $s_{i}$. Therefore, the sum of the entries in the $i$th row of $\textbf{N}$ is the expected times in any of the transient states for a given starting state $s_i$, i.e., the expected time required before the follower transfers to candidate state.
Denoting $s_{1}$ as the initial state, we obtain the proposition. \ \ \ \ \ \ \ \ \ \ \ \ \ \ \ \ \ \ \ \ \ \ \ \ \ \ \ \ \ \ \ \ \ \ \ \ \ \ \ \ \ \ \ \ \ \ \ \ \ \ \ \ \ \ \ \ \ \ \ \ \ \ \ \  \ \ \ \ \ \ \ \ \ \ \ \ \ \ \ \ \ \ \ \ \ \ \ \ \ \ \ \ \ \ \ \ \ \ \ \ \ \ \ \ \ \ \ \ \ \ \ \ $\blacksquare$

\textbf{Property 3:} The average interval of the received heartbeats for a follower in one term is given by
\begin{equation}
t_{in}= \frac{\sum \limits_{j=1}^{t}n_{1j}}{n_{11}}.
\label{equ_2-22}
\end{equation}

\section{simulation results}
In this section, we first validate the efficiency of the analysis model, and then, we investigate the impacts of the parameters (such as packet loss rate, election timeout period, and network size) on availability and network split probability in more details.

To validate the model, we have compared its results with those obtained with the Raft simulator from \cite{IEEEhowto:Ongaro}.
In simulations, each message was assigned a latency chosen randomly from the uniform range of $[0.5, 10]$ms, and the interval between two heartbeats $h=50$ ms.
Furthermore, CPU time should be short relative to network latency, and the speed of writing to disk does not play a significant role anyhow.

\begin{figure}[h]
\begin{centering}
\includegraphics[scale=0.5]{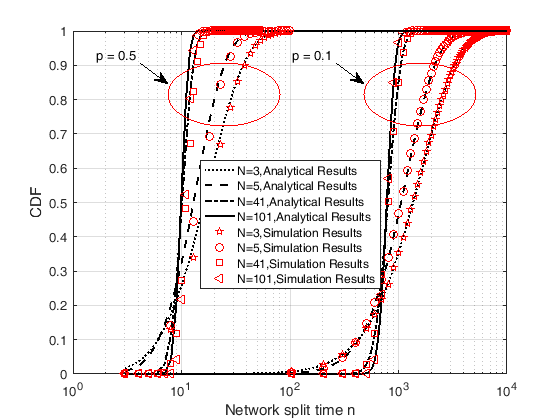}
\end{centering}
\caption{CDF of network split time, $K=3$}
\label{fig_sim1}
\end{figure}

\begin{figure}[h]
\begin{centering}
\includegraphics[scale=0.5]{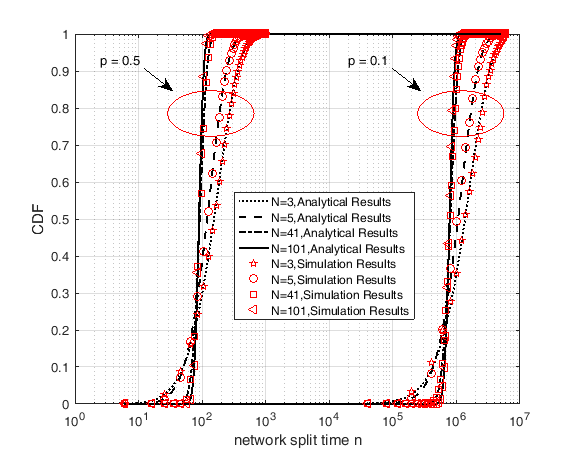}
\end{centering}
\caption{CDF of network split time, $K=6$}
\label{fig_sim2}
\end{figure}
%
%
%

\begin{figure}[h]
\begin{centering}
\includegraphics[scale=0.5]{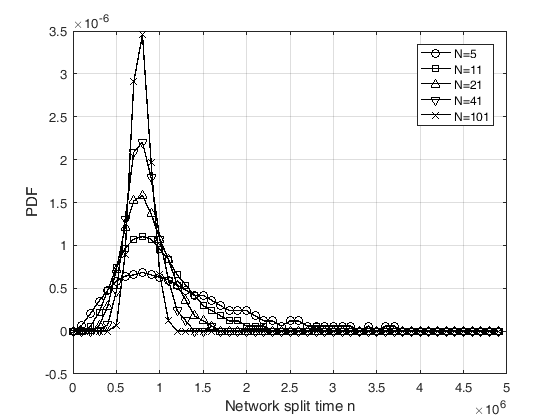}
\end{centering}
\caption{PDF of network split time $K=6$, $p=0.1$}
\label{fig_sim8}
\end{figure}

\begin{figure}[h]
\begin{centering}
\includegraphics[scale=0.5]{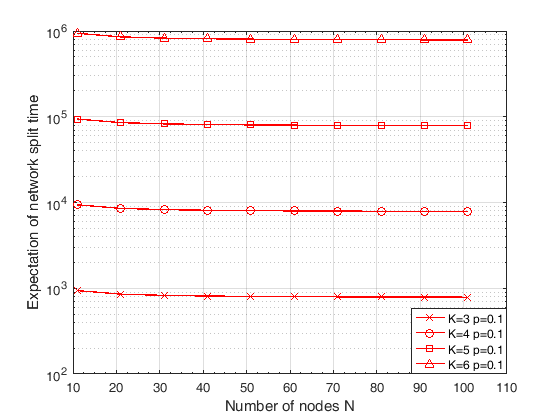}
\end{centering}
\caption{Expectation of network split time given different network sizes}
\label{fig_sim9}
\end{figure}

\begin{figure}[h]
\begin{centering}
\includegraphics[scale=0.5]{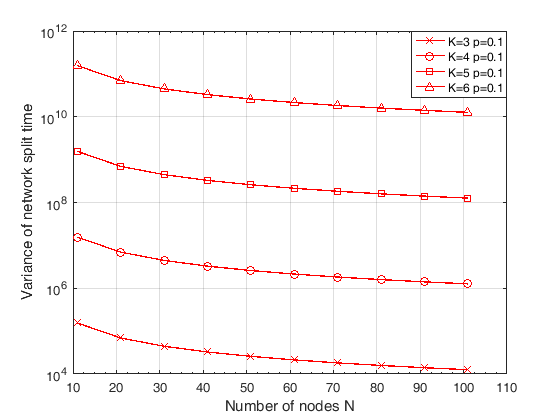}
\end{centering}
\caption{Variance of network split time given different network sizes}
\label{fig_sim10}
\end{figure}

\begin{figure}[h]
\begin{centering}
\includegraphics[scale=0.5]{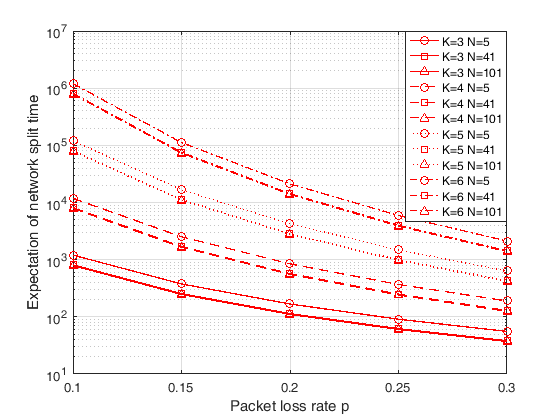}
\end{centering}
\caption{Expectation of network split time given different packet loss rates}
\label{fig_sim12}
\end{figure}

\begin{figure}[h]
\begin{centering}
\includegraphics[scale=0.5]{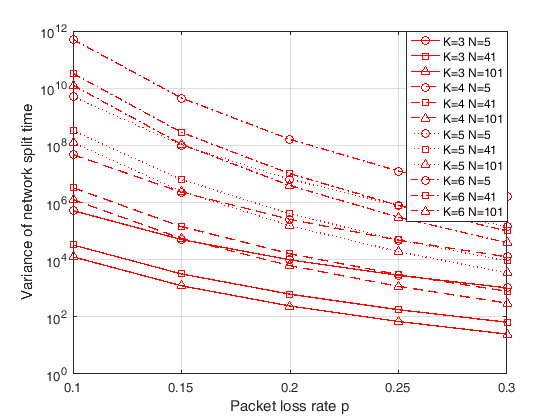}
\end{centering}
\caption{Variance of network split time given different packet loss rates}
\label{fig_sim13}
\end{figure}

\begin{figure}[h]
\begin{centering}
\includegraphics[scale=0.5]{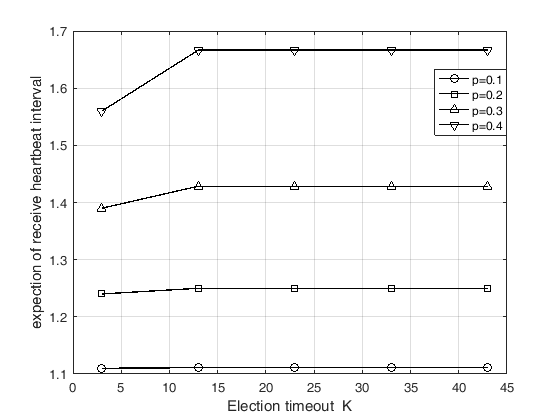}
\end{centering}
\caption{Average interval to receive a heartbeat for a follower}
\label{fig_sim14}
\end{figure}

Figs. \ref{fig_sim1}-\ref{fig_sim2} plot the cumulative
distribution function (CDF) of the number of heartbeats for a network to split. Each CDF summarizes $10,000$ simulated trials. The analytical results are calculated based on Eq. (\ref{equ_2-19}). Given the value of $K$, network size $N$ and packet loss rate $p$, Figs. \ref{fig_sim1}-\ref{fig_sim2} show that analytical results match well with the simulation results. Therefore, one can detect when the network is abnormal by comparing with the reference value given by the analytical model.

From Figs. 2 and 3, we observe that (i) network split probability highly depends on the packet loss rate $p$ and the election timeout value $K$. As one expected, as $p$ or $K$ increases, for the same size of network, the network split probability decreases. (ii) As the network size increases, the CDF curves become steeper. Given $p$ and $K$, when $n$ is small, $p_{n}$ decreases with when $N$ increases. This can be observed from $p_{n}$ in Eq. (\ref{equ_2-19}). To further understand the second point, we show the PDF of network split time. Fig. \ref{fig_sim8} shows the PDF of network split time for different network sizes. We observe that large network has smaller split probability than the one in small network at the beginning of running time. When running time increases over certain point, the network split probability increases with the size of network. Because the follower's probability of transition to candidate is small at the beginning of running time, the probability that more than half of all nodes become candidates is lower for larger networks. When the follower's probability of transition to candidate gets greater with the running time, the network split probability increases with the network size.

The analytical model given in Section III is convenient to determine the probability of network split time. Based on the proposed model, we exploit the impacts of the parameters (such as packet loss rate, election timeout period, and network size) on availability and network split performance in more details. Fig. \ref{fig_sim9} and Fig. \ref{fig_sim10} describe the expectation and variance of network split time for different network sizes, respectively. The results show that the variance of network split time for large network is smaller than that for small network. However, the expectation of network split time for larger network is very close to that for small network. Therefore a larger network has better stability in terms of network split time.


Fig. \ref{fig_sim12} and Fig. \ref{fig_sim13} show the expectation and variance of network split time in different packet loss rates, respectively.
As Fig. \ref{fig_sim12} shows, given the packet loss rate $p=0.1$ and $N=5$, the expectation of split time is about 1,000 and 10,000 when $K=3$ and $K=4$, respectively. This result means that the network's stable time is prolonged 10 times by adding one more heartbeat. Given the packet loss rate $p=0.3$ and $N=5$, the expectation of split time is about 50 and 110 when $K=3$ and $K=4$, respectively. The expectation of network split time is prolonged 1 times by adding one more heartbeat. Increasing election timeout is helpful to lower the network split probability caused by packet loss, especially under smaller packet loss rate.

Fig. \ref{fig_sim14} plots the average interval of receiving one heartbeat for one follower per term. The results show that packet loss rate has significant impact on receiver heartbeat interval. On the other hand, election timeout period has insignificant impact on receiver heartbeat interval. Especially, the interval of receiving one heartbeat increases with election timeout period but goes to a constant.

\section{Conclusion}
Consensus is the basis for blockchains. Raft is a well adopted consensus algorithm for private blockchains. In this paper, an analytical model for Raft consensus algorithm is proposed. The analytical model given in this paper is very convenient to obtain the network split probability, which provides guidance on how to determinate the parameters such as election timeout. Furthermore, the analytical model is able to monitor the condition of network and detect the abnormal condition by providing reference value of network performance. The simulation results match the analytical results well. Using the proposed model, we have shown the parameters'(such as packet loss rate, election timeout and network size) impacts on availability. Increasing election timeout is helpful to lower the network split probability caused by packet loss. We also observe that larger network has smaller split probability than the one in small network at the beginning of running time and more focused splitting time.

\appendix
(a) The absolute values of eigenvalue of $\textbf{Q}$ are all strictly less than 1.

Based on Gershgorin circle theorem \cite{IEEEhowto:https}, for $K\times K$ matrix $\textbf{Q}$, all eigenvalue satisfy

\begin{equation}
|a_{k}-q_{ii}|\leq \sum_{j\neq i}|q_{ij}|,  \ \ \  i, k=1,2,\cdots,K,   \label{equ_2-200}
\end{equation}
where $q_{ij}$ is the $(i,j)$th element of $\textbf{Q}$.

Due to $\sum \limits_{j=1}^{K}q_{ij}\leq 1$ and $q_{ij}\geq 0$,
\begin{eqnarray}
\nonumber
-\sum_{j\neq i}q_{ij}\leq a_{k}-q_{ii} \leq \sum_{j\neq i}q_{ij} \ \ \  \ \ \ \ \ \ \ \ \ \ \  \ \\
-\sum_{j=i}^{K}q_{ij}+2q_{ii}\leq a_{k} \leq \sum_{j=i}^{K}q_{ij}, \ \ \ \ \ \forall i, k=1,2,\cdots,K.
\end{eqnarray}

Since $\mathbf{R}$ in Eq. (\ref{equ_2-300}) is not zero matrix, there exists one row of $\mathbf{Q}$ such that $\sum \limits_{j=1}^{K}q_{ij}< 1$. Thus,

\begin{equation}
-1 < a_{k} < 1,  \ \ \  k=1,2,\cdots,K.   \label{equ_2-200}
\end{equation}

That means $\mid a_{k} \mid < 1$.

(b) Proof of $\lim \limits_{n\rightarrow \infty}\textbf{Q}^{n}= \textbf{0}$

Note that $\mathbf{Q}=\mathbf{U}\mathbf{\Lambda} \mathbf{U}^{\texttt{H}}$, where $\mathbf{U}^{\texttt{H}}\mathbf{U}=\mathbf{I}$ and
\begin{equation}
\mathbf{\Lambda}=
 \left [
 \begin{matrix}
  a_{1} & & \\
  &\ddots & \\
  & & a_{t}
  \end{matrix}
  \right ].\label{equ_2-202}
\end{equation}

Since $|a_{i}|<1$, we have $\lim \limits_{n\rightarrow \infty} \mathbf{\Lambda}^{n}= \textbf{0}$.

Furthermore, since $\mathbf{Q}^{n}=\mathbf{U}\mathbf{\Lambda}^{n} \mathbf{U}^{\texttt{H}}$,

$\therefore \lim \limits_{n\rightarrow \infty}\textbf{Q}^{n}= \textbf{0}$.


\end{document}